\documentclass[aip,reprint]{revtex4-1}

\usepackage{graphicx}
\usepackage{dcolumn}
\usepackage{bm}

\usepackage[utf8]{inputenc}
\usepackage[T1]{fontenc}
\usepackage{mathptmx}
\DeclareMathAlphabet{\altmathcal}{OMS}{cmsy}{m}{n}

\usepackage{hyperref}
\hypersetup{
colorlinks   = true,
citecolor    = blue,
linkcolor    = blue}

\renewcommand{\v}[1]{\ensuremath{\bm{#1}}} 
 
\newcommand{\uv}[1]{\hat{\ensuremath{\bm{#1}}}} 
\newcommand{\pd}[2]{\frac{\partial #1}{\partial #2}} 
 
\let\baraccent=\= 
\renewcommand{\=}[1]{\stackrel{#1}{=}} 

\newcommand{\zintm}{\int_{z_{\min}}^z}
\newcommand{\zintmm}{\int_{z_{\min}}^{z_{\max}}}

\newcommand{\bhat}{\uv{b}}

\newcommand{\Gbar}{\bar{\Gamma}}
\newcommand{\nbar}{\bar{n}}
\newcommand{\cs}{c_s}
\newcommand{\upars}{u_{\parallel s}}
\newcommand{\upari}{u_{\parallel i}}

\newcommand{\ubar}{\bar{u}}

\newcommand{\zmin}{z_{\min}}
\newcommand{\zmax}{z_{\max}}

\newcommand{\vpar}{v_{\parallel}}

\newcommand{\ExB}{$\v{E} \times \v{B}$}

\newcommand{\gkyl}{{\tt Gkeyll}}
\newcommand{\etal}{\emph{et al.}}
\newcommand{\vparmax}{v_{\parallel,\max}}

\draft 

\begin{document}


\title{Investigating shear flow through continuum gyrokinetic simulations of limiter biasing in the Texas Helimak}

\author{T. N. Bernard}
    \email{bernardt@fusion.gat.com}
    \affiliation{General Atomics, San Diego, CA 92121, USA}
    \affiliation{Oak Ridge Associated Universities, Oak Ridge, TN 37830, USA}
    \affiliation{Institute for Fusion Studies, University of Texas at Austin, Austin, TX 78712, USA}
\author{T. Stoltzfus-Dueck}
    \affiliation{Princeton Plasma Physics Laboratory, Princeton, NJ 08543, USA}
\author{K.~W. Gentle}
    \affiliation{Institute for Fusion Studies, University of Texas at Austin, Austin, TX 78712, USA}
\author{A.~Hakim}
    \affiliation{Princeton Plasma Physics Laboratory, Princeton, NJ 08543, USA}
\author{G.~W. Hammett}
    \affiliation{Princeton Plasma Physics Laboratory, Princeton, NJ 08543, USA}
\author{E.~L. Shi}
    \affiliation{Lawrence Livermore National Laboratory, Livermore, CA 94550, USA}
    \affiliation{Department of Astrophysical Sciences, Princeton University, Princeton, NJ 08544, USA}


\date{\today}

\begin{abstract}
Previous limiter-biasing experiments on the Texas Helimak, a simple magnetized torus, have been inconclusive on the effect of flow shear on turbulence levels. To investigate this, the first gyrokinetic simulations of limiter biasing in the Helimak using the plasma physics code \gkyl\ have been carried out, and results are presented here. For the scenarios considered, turbulence is mostly driven by the interchange instability, which depends on gradients of equilibrium density profiles. An analysis of both experimental and simulation data demonstrates that shear rates are mostly less than than local linear growth rates, and not all requirements for shear stabilization are met. Rather, the mostly vertical shear flow has an important effect on bulk transport and experimental equilibrium density profiles, and changes to the gradients correspond to changes in turbulence levels.
\end{abstract}


\maketitle 

\section{Introduction} \label{sec:intro}
Since the discovery of the H-mode in the ASDEX tokamak \citep{wagner1982regime} and in other devices thereafter, there has been significant interest in understanding the conditions that govern the transition to this improved confinement regime. H-mode is characterized by increased core energy, steep gradients in the density and temperature profiles near the edge of the plasma, and a reduction of radial turbulent transport across the separatrix, or last closed magnetic flux surface. A sheared, poloidal flow in the plasma edge, generated by radial electric fields, has been observed to correlate with this reduction in turbulence,\citep{Burrell_1989} and the role of \ExB\ flow shear in turbulence suppression has been studied extensively. \citep{biglari1990influence,burrell1997,terry2000suppression} In strongly magnetized plasmas, shear may break up turbulent structures and, hence, reduce turbulence levels, provided certain requirements are met: (1) the shear rate is greater than the turbulence decorrelation rate, (2) turbulent structures remain in the region of shear longer than the decorrelation time, and (3) the flow itself is stable, i.e. not Kelvin-Helmholtz unstable. \citep{terry2000suppression} 
\begin{figure}
    \centering
    \includegraphics[width=0.45\textwidth]{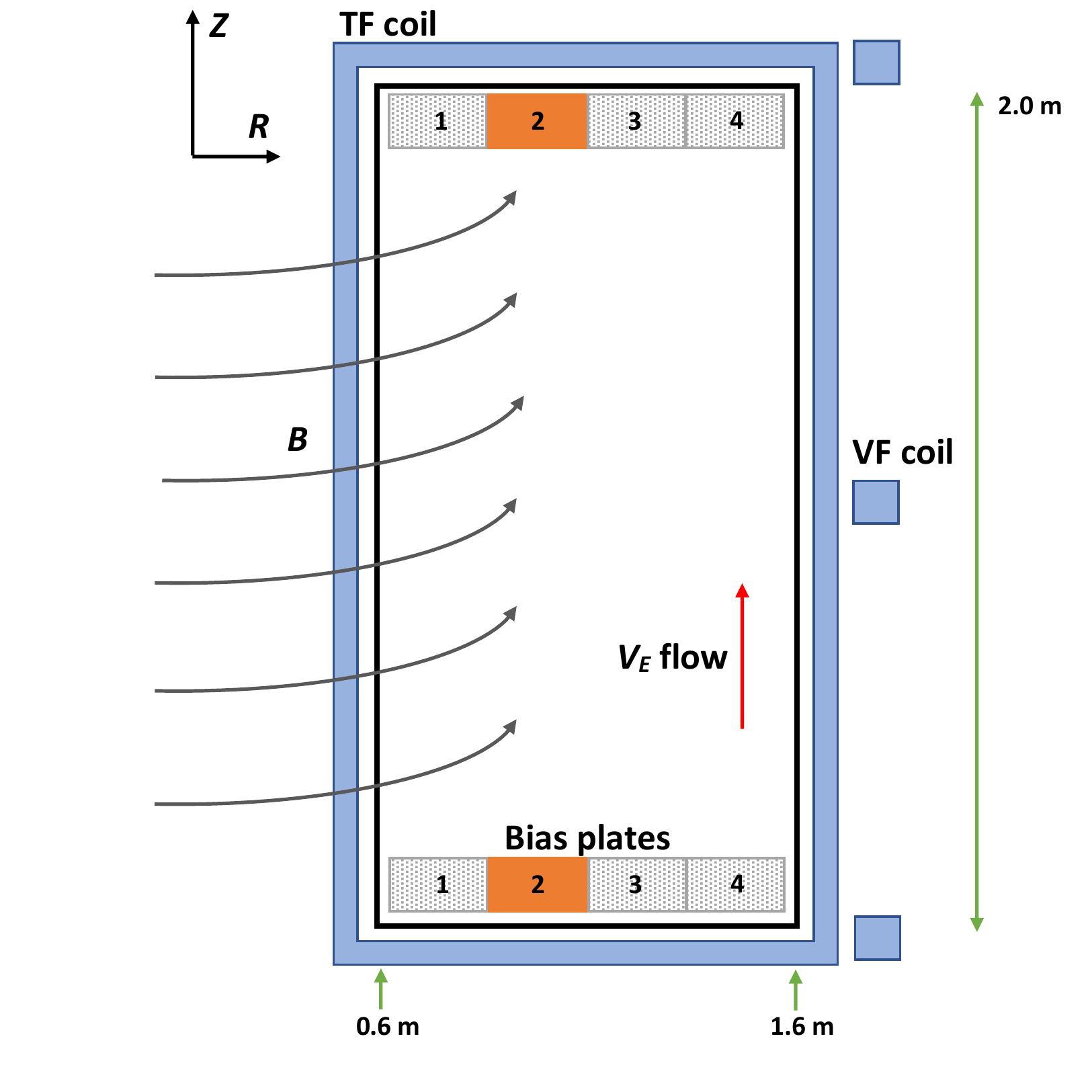}
    \caption{Schematic of the Helimak toroidal cross section with the vacuum vessel outlined in black.  In scenarios considered here, the second limiter plates (solid orange) were biased to a given voltage, while the others (dotted gray) were grounded. An identical set of plates appears at a toroidal location $180^\circ$ apart from this image.}
    \label{fig:helimak-cs}
\end{figure}

The effect of flow shear on turbulence levels has been studied experimentally in more basic, open-field-line plasma devices such as LAPD,\citep{zhou2012} TORPEX, \citep{theiler2012torpex} and the Texas Helimak \citep{gentle2010comparison,gentle2014turbulence} by applying a bias voltage to plasma limiters, which drives a sheared \ExB\ flow. This provides more direct control over the shear than is typically available in the edge of a tokamak. Simple magnetized torus experiments, like TORPEX and the Helimak, have helical, open field lines and are good approximations to a tokamak scrape off layer (SOL), with the radial and toroidal directions analogous to those of a tokamak and the vertical $Z$ direction analogous to the poloidal direction. We focus this study on the Helimak, whose field lines terminate on conducting plates at the top and bottom of the vacuum vessel, pictured in Fig.~\ref{fig:helimak-cs}. These are arranged in four sets of four plates, with two sets on the top and two sets on the bottom, located $180^\circ$ apart toroidally. Each plate is approximately 0.2 m in width and 0.2 m in height, oriented vertically such that the toroidal direction is normal to the plate. A bias voltage can be applied to the same plate in each of the four sets, such as the second set of plates from the inner Helimak wall, leaving the other three grounded, to modify the velocity shear. A range of -40 V to 15 V has been applied to bias plates. The plasma response to sheath physics at the plates sets up a radially varying electric potential, resulting in a sheared \ExB\ flow that is mostly vertical since the toroidal component of the magnetic field is much stronger than the vertical component ($B_Z/B_\phi \lesssim 0.05$). This vertical flow has been observed experimentally via spectroscopic measurements of ion flow.\cite{gentle2010comparison,gentle2014turbulence} Since ions are relatively cold, other particle drifts are negligible, and the \ExB\ drift dominates.

The magnetic field line connection length can be scanned by changing resistance through the vertical field coils. We consider a low connection-length scenario ($L_c(R_0) \approx 40$ m), with $k_\parallel \approx 0$, which is dominated by interchange turbulence.\cite{Williams2017thesis} This instability occurs in the ``bad-curvature region," where $\kappa \cdot \nabla_x n > 0$; $\kappa$ is the inward-pointing radius of curvature and $\nabla_x n$ is the radial gradient of the equilibrium density profile. This region begins around $R = 1$ m in the device and extends to larger radii. To test the effect of shear flow on interchange turbulence, a bias potential can be applied on a plate in the bad-curvature region, which corresponds to plates 2, 3, or 4. However, for the low connection length considered here, some field lines at plates 3 and 4 terminate on the grounded vessel wall rather than the plates, due to the radial variation in connection length. Thus, the second conducting plates (highlighted in Fig.~\ref{fig:helimak-cs}) were biased in the data presented here. It has been demonstrated that higher connection-length scenarios, with finite $k_\parallel$, are dominated by the drift-interchange instability and analysis of these is beyond the scope of this work.

An array of Langmuir probes is located on the bias plates to calculate equilibrium profiles and turbulence levels across the full radial extent of the device and along a smaller range in $Z$. \ExB\ flow has previously been calculated using spectroscopic measurements, which is described in more detail in Ref.~\citenum{gentle2014turbulence}. 

Standard theories of shear stabilization have not been sufficient to explain observations from previous limiter-biasing experiments on the Helimak. According to Ref.~\citenum{gentle2014turbulence}, shear structures remain in the region longer than linear growth times, satistfying one of the requirements for shear stabilization. The authors also state that shear rates (Fig.~15) are equal to or larger than linear growth rates, though they do not include data for the latter. A comparison of reduction in turbulence fluctuation levels vs.~shear was plotted using data from numerous experiments (Fig.~17 in Ref.~\citenum{gentle2014turbulence}) and showed no clear correlation between the two quantities.

In order to better understand those observations, we have performed continuum gyrokinetic simulations of limiter-bias scenarios in the Helimak with the plasma physics code \gkyl, building upon simulations of limiter biasing in LAPD by Shi\cite{Shi2017thesis} and simulations of a grounded (no limiter biasing) argon discharge in the Helimak by Bernard \etal \cite{bernard2019} Though fluid simulations of biasing in the Helimak have been performed and resulting profiles approach experimental values, \citep{li2011turbulence} kinetic models are necessary to reproduce non-Maxwellian features in the particle distribution functions. For example, it has been shown that there are small but non-negligible\ kinetic sheath effects in the Helimak.\cite{bernard2019} Furthermore, our model introduces \ExB\ velocity shear self-consistently via the plasma response to boundary conditions in the parallel direction. This produces a radial gradient in the electrostatic potential, resulting in the radial electric field that gives rise to a sheared \ExB\ flow. 

The paper is organized as follows: In Sec.~\ref{sec:modeleqns}, we begin with an explanation of the model equations and simulation set-up. In Sec.~\ref{sec:results}, we present the results of limiter-biasing simulations of the Texas Helimak and make comparisons with the grounded simulation and experimental data. We find fair agreement and similar trends with experimental equilibrium profiles and also describe the effect of shear flow on density profiles. For scenarios considered here, most shear rates are less than local linear growth rates, and requirements for shear stabilization are not met. We give an explanation for how shear flow affects turbulence by changing gradients of density profiles. In Sec.~\ref{sec:conc}, we conclude with a summary of our findings.

\section{Model equations} \label{sec:modeleqns}
The \gkyl\ code has previously been used to model helical, open-field-line plasmas with the continuum gyrokinetic solver,\cite{Shi2017thesis,shi2019full,bernard2019} and we briefly describe the model and parameters used in the code to produce the simulations presented below. \gkyl\ uses the nodal discontinuous Galerkin computational method for the spatial discretization and an explicit third-order Runge--Kutta method to discretize in time. The full-$f$ gyrokinetic equation in the long-wavelength, zero-Larmor-radius limit is used to evolve the gyrocenter distribution function $f_s(\bm{R}, v_\parallel, \mu,t)$
\begin{equation}
\label{eq:gk-eqn}
\frac{\partial \altmathcal{J} f_s}{\partial t} + \nabla \cdot (\altmathcal{J} \dot{\bm{R}} f_s) + \frac{\partial}{\partial v_\parallel} (\altmathcal{J} \dot{v}_\parallel f_s )
=\altmathcal{J} C[f_s] + \altmathcal{J} S_s,
\end{equation}
where $C[f_s]$ represents the effect of collisions and $S_s$ is a source term. The Jacobian is $\altmathcal{J} = B^*_\parallel$, where $B_\parallel^* = \bm{b} \cdot \bm{B}_\parallel^*$, and $\bm{B}_\parallel^* = \bm{B} +(Bv_\parallel/\Omega_s)\nabla \times \bm{b}$. The phase-space advection velocities $\dot{\bm{R}} = \{\bm{R},H\}$ and $\dot{v}_\parallel = \{v_\parallel,H\}$ are defined in terms of the Poisson bracket
\begin{eqnarray}
\{F,G\} = \frac{\bm{B}^*}{m_s B_\parallel^*} \cdot \left( \nabla F \frac{\partial G}{\partial v_\parallel} - \frac{\partial F}{\partial v_\parallel} \nabla G \right) \nonumber \\
- \frac{1}{q_s B_\parallel^*} \bm{b} \cdot \nabla F \times \nabla G.
\label{eq:pb-eqn}
\end{eqnarray}
In the long-wavelength limit, the gyrocenter Hamiltonian is
\begin{equation}
H_s = \frac{1}{2}mv_\parallel^2 + \mu B + q_s \phi.
\end{equation}
The long-wavelength gyrokinetic Poisson equation, which is a statement of quasineutrality, is used to evolve the electrostatic potential
\begin{equation}
 -\nabla \cdot \left( \frac{n_{i0}^g q_i^2 \rho_{\mathrm{s}0}^2}{T_{e0}} \nabla_\perp \phi \right) = \sigma_g
 = q_i n_i^g(\bm{R},t) - e n_e(\bm{R},t),
 \label{eq:poisson}
 \end{equation}
where $\rho_{\mathrm{s}0} = c_{\mathrm{s}0}/\Omega_i$ is the ion sound gyroradius and $c_{\mathrm{s}0} = \sqrt{T_{e0}/m_i}$ is the ion sound speed. According to spectroscopic measurements of the Helimak, the argon ions have a charge state of $Z=1$, and thus we assume $q_i = e$. In the long-wavelength limit, gyro-averaging is neglected, and this assumption will be relaxed in the future.

\gkyl\ employs a nonorthogonal field-line-following coordinate system,\cite{beer1995field,hammett1993developments,scott1998global} where $z$ is the distance along the field, $x$ is the radial coordinate, and $y$ is the bi-normal coordinate. Geometric factors due to the cylindrical coordinate system are only retained in $\bm{B}^*$ and neglected elsewhere. Like most flux-tube codes, we assume that perpendicular gradients are much stronger than parallel gradients and make the following approximation:
\begin{eqnarray*}
(\nabla \times \bm{b})\cdot \nabla &=& [(\nabla \times \bm{b}) \cdot \nabla y]\pd{}{y} + [(\nabla \times \bm{b}) \cdot \nabla z]\pd{}{z} \\
&\approx& [(\nabla \times \bm{b}) \cdot \nabla y]\pd{}{y}, \label{eq:}
\end{eqnarray*}
where $\nabla y$ is the reciprocal-basis vector of the curvilinear coordinate system. Field-line curvature enters through this term as $(\nabla \times \bm{b}) \cdot \nabla y = -1/x $. Magnetic shear is not included in the model, and since the field strength is dominated by the vacuum toroidal field, we assume $B(x)= B_0(R_0/x)$. 

Limiter biasing is modeled via the ``conducting-sheath'' boundary conditions in the $z$ direction, which is parallel to the magnetic field. For now, our model assumes that field lines intersect the plates perpendicularly, though in reality they intersect at a weakly non-normal angle that depends on field-line pitch angle, which can be varied in the experiment. We use Eq.~\ref{eq:poisson} to solve for the sheath potential $\phi_{\mathrm{sh}}$ at this boundary. Because our gyrokinetic model assumes quasineutrality (Eq.~\ref{eq:poisson}), we do not resolve the Debye sheath, in which $n_i > n_e$. The wall is assumed to be just outside the simulation domain with a potential $\phi_\mathrm{w}$. The cut-off parallel velocity for electrons is $v_c = \sqrt{\max{(2e(\phi_{\rm sh} - \phi_{\rm w})/m_e,0)}}$, determining which electrons are reflected at the wall ($\vpar < v_c$) and which leave the domain ($v_\parallel > v_c$). For the grounded scenario, $\phi_\mathrm{w}=0$. To model limiter biasing, we specified $\phi_\mathrm{w}=V_b$ at a radial interval to simulate a bias voltage on the second-from-the-inner-wall conducting plates in the Helimak:
\begin{equation}
\label{eq:bc-bias}
\phi_\mathrm{w}(x,y) = \left\{
    \begin{array}{cc}
        V_b & x_L \leq x \leq x_R  \\
        0 & \mathrm{else},
    \end{array} \right.
\end{equation}
where $x_L$ and $x_R$ represent the radial extent of the biased plate. Thus, the plasma potential in this model evolves self-consistently via the gyrokinetic Poisson equation and the parallel boundary conditions. We maintained the Dirichlet boundary conditions in $x$ for the potential ($\phi = 0$) and periodic boundary conditions in $y$ for $f$ and $\phi$. 

To simulate the biased discharges, we restarted two simulations from the grounded case at 4 ms, which is approximately one ion transit time, with different bias voltages and ran to 16 ms. A bias voltage was applied to the second plate (highlighted in Fig.~\ref{fig:helimak-cs}) with $x_L = 0.86$ m and $x_R = 1.06$ m. We simulated the most negative bias voltage, ${V_b = -40}$~V, and the most positive, ${V_b = 10}$ V, used in recent experiments. (For positive bias, the back side of the plates draw significant current, and the experimental apparatus has a current-limiting resistor that reduces the voltage actually applied to the plates. Therefore, an engineering value of +15 V is given to control the power supply, but, due to resistance, only approximately +10 V is applied to the plates.) All other parameters are the same as in the grounded simulation to facilitate direct comparison, including a magnetic field line connection length ${L_z = 40}$ m. Since simulations do not include magnetic shear, we compared simulation results with experimental data from discharges with a similar connection length at the radial center of the vacuum chamber. This corresponded to ${B_Z \simeq 0.005}$~T and ${B_Z/B_\varphi \simeq 0.05}$. Simulation equilibrium profiles and other statistics were calculated over a 10 to 16 ms window. Simulation parameters are summarized in Table~\ref{tab:table1}.
\begin{table}
\caption{\label{tab:table1}Summary of Helimak simulation parameters}
\begin{ruledtabular}
\begin{tabular}{ll}
Vacuum vessel height           & $H = 2.0 $ m \\
Background magnetic field & $B_0 =  0.1$ T \\
Background density & $n_0 = 10^{16}$ m$^{-3}$\\
Background electron temperature & $T_{e0} = 10$ eV\\
Background ion temperature & $T_{i0} = 1$ eV\\
Electron gyroradius & $\rho_e = 1.02 \times 10^{-2}$ m \\
Electron mean free path & $\lambda_{ee}$ = 98.4 m \\
Ion sound gyroradius & $\rho_{\mathrm{s}0} = 2.04 \times 10^{-2}$ m \\
Ion sound speed & $c_\mathrm{s} = 4.90 \times 10^{3}$ m/s \\
Ion mean free path & $\lambda_{ii}$ = 1.39 m \\
Limiter bias voltage & $V_b = 0, 10, -40$ V
\end{tabular}
\end{ruledtabular}
\end{table}

\section{Comparison with Experiment} \label{sec:results}
\begin{figure}
    \centering
    \includegraphics[width=.5\textwidth]{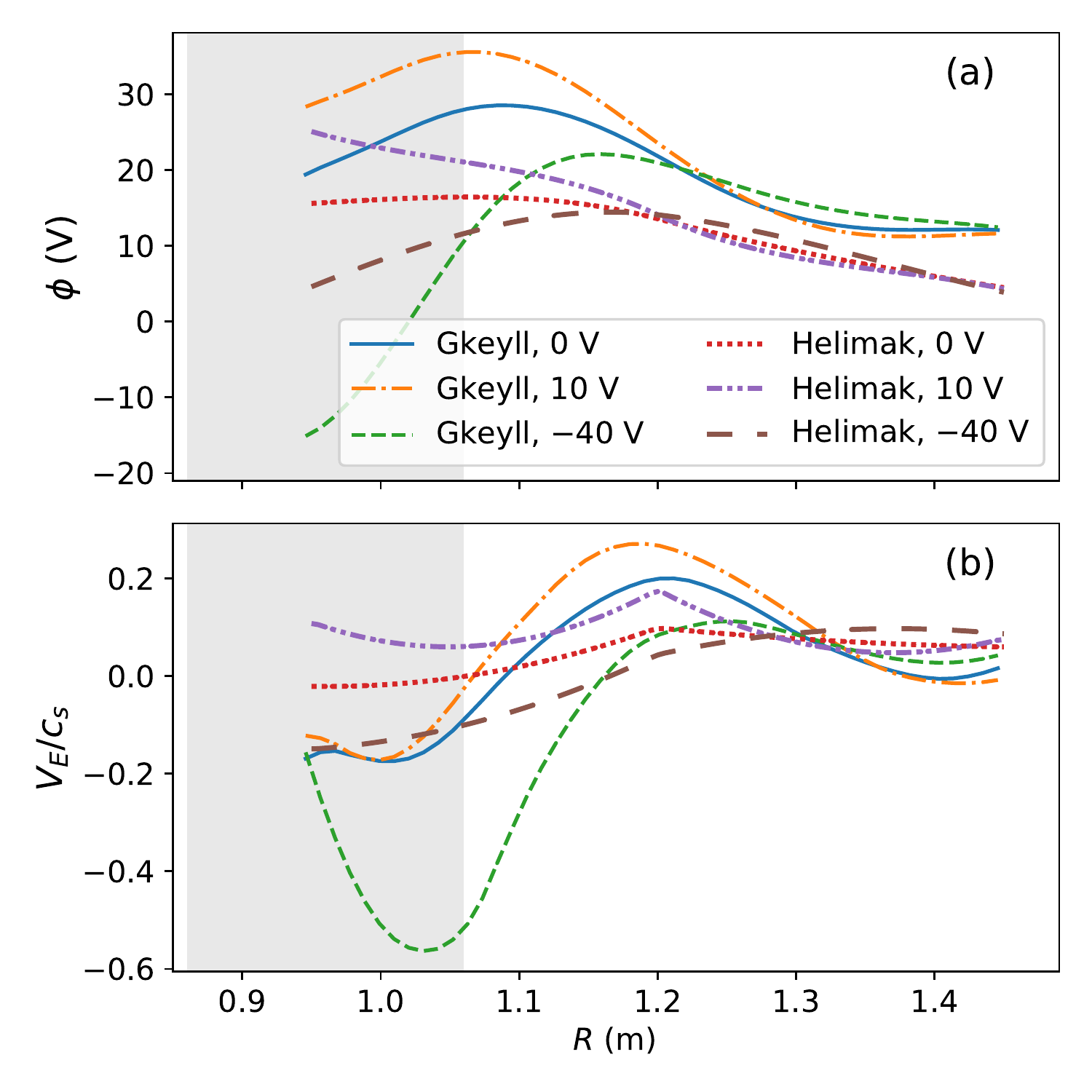}
    \caption[Comparison of the electrostatic potential and \ExB\ flow for grounded and limiter-biased scenarios of both simulation and experiment]{Plot (a) compares the electrostatic potential profiles and (b) compares \ExB\ flow for grounded and limiter-biased scenarios of both simulation and experiment. The experimental plasma potential is measured with a baffled probe. The \ExB\ flow is calculated from the plasma potential as $V_E = -(d\phi_0/dx)/B_0$, where $x$ is the radial coordinate.}
    \label{fig:phi-ve}
\end{figure}
A grounded, a positively biased, and a negatively biased experiment were carried out on the Texas Helimak for comparison with simulations. A resistance of $2.7$~$\Omega$ in the vertical field coils was used, which approximately corresponds to a connection length of 40 m at the radial center $R_0=1.1$ m. The experimental connection length varies with radius, and this is not included in our model, which currently neglects magnetic shear. Within this section we separately consider the effect of shear flow on equilibrium profiles and the effect of shear flow on turbulence levels, as well comparisons of other measurements, such as temperature and potential fluctuations and parallel heat flux to the limiter. 

\subsection{Comparison of equilibrium profiles}
The effect of the limiter biasing on the equilibrium plasma potential profile is clearly seen in Fig.~\ref{fig:phi-ve}a. A baffled probe is used to directly measure the experimental plasma potential equilibrium profile \citep{demidov2010} and these values may have a 20\% systematic error. An increase or decrease in the plasma profile is observed corresponding to a positive or negative bias voltage in both simulation and experimental data. Also, shown with these plots is the \ExB\ flow, which was calculated from the equilibrium plasma potential profile as $V_E = -(d\phi_0/dx)/B_0$, where $x$ is the radial coordinate, and normalized to the ion sound speed. Direct spectroscopic measurements of experimental flow were not available for these discharges, and previous measurements of the vertical flow\cite{gentle2010comparison,gentle2014turbulence} are about 2--4 times greater in magnitude than the $V_E$ flow calculated from the electrostatic potential, shown in Fig.~\ref{fig:phi-ve}b. 

By contrast, \ExB\ flow in the simulation is mostly in the $y$-direction, which corresponds to the toroidal direction in the experiment. (See Fig.~2 in Ref.~\citenum{bernard2019} for a more detailed explanation of the mapping from the nonorthogonal coordinate system to cylindrical coordinates.) This is due to the $k_\perp \gg k_\parallel$ assumption made in most flux-tube codes. Thus, the advection due to a radial electric field is approximated as 
\begin{equation}
V_E \cdot \nabla 
= V_E \cdot \nabla y \, \pd{}{y}
+ V_E \cdot \nabla z \, \pd{}{z} \approx V_E \cdot \pd{}{y}. \label{eq:advec}
\end{equation} 
In the nonorthogonal field-aligned coordinate system, $\nabla z \sim L_z/H \approx B_\phi/B_Z$, where $L_z$ is the magnetic field line connection length and $H$ is the height of the vacuum vessel. A more appropriate condition would be $k_\perp \gg k_\parallel B_\phi/B_Z $. This is satisfied for the parameters of this simulation since $k_\parallel \approx 0$.\cite{bernard2019} We would expect the vertical $V_E$ flow to have little impact on turbulence levels or statistics. However, if simulations with longer connection lengths were carried out, $B_\phi/B_Z$ would increase and $k_\parallel$ would become finite.\cite{poli2008transition,ricci2009transport,ricci2010turbulence} Hence, this relation may no longer be satisfied. 

\begin{figure}
    \centering
    \includegraphics[width=.5\textwidth]{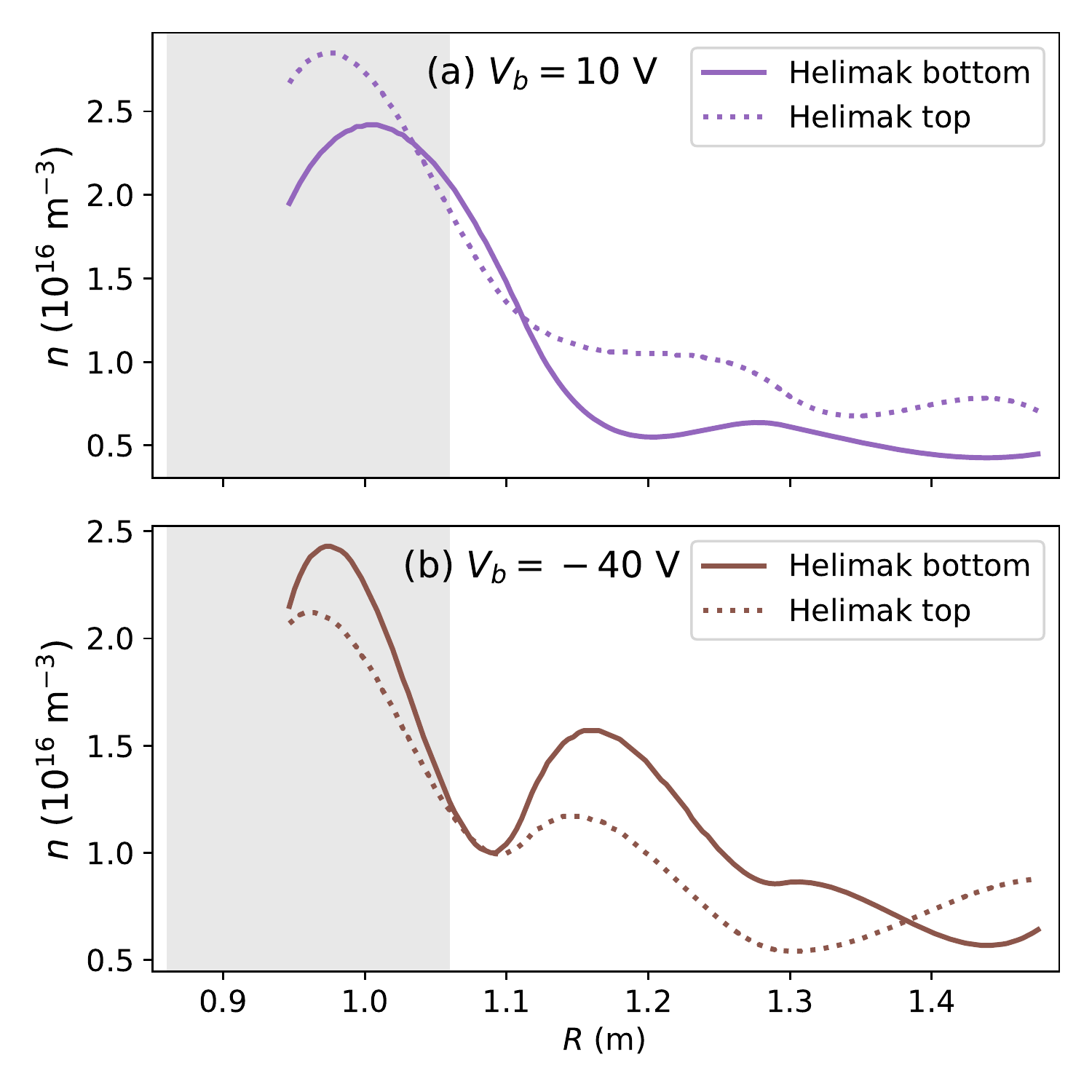}
    \caption{Comparison of experimental equilibrium density profiles measured from the bottom and top of the Helimak vessel for cases with (a) positive limiter biasing and (b) negative limiter biasing.}
    \label{fig:n-TB}
\end{figure}

The assumption in equation \ref{eq:advec} does not account for the effect of the vertical component of $V_E$ on bulk transport and equilibrium profiles. We provide a estimate for this effect in Appendix \ref{app} using a simplified 1D transport model. We assume a ``moderate'' vertical \ExB\ flow that is the same order of magnitude of the projection of the parallel flow on the vertical coordinate. We conclude that equilibrium density profile values should be larger ``downstream," that is, at the end of the device towards which the \ExB\ flow is directed. Qualitative agreement with this is observed in Fig.~\ref{fig:n-TB}, which shows top and bottom density profiles from the (a) positively biased case and (b) negatively biased case. Looking at the experimental $V_E$ flow for the positively biased case in Fig.~\ref{fig:phi-ve}b, one can see that the flow is positive and hence the top of the Helimak is the downstream side. The top profile is in most regions higher than the bottom profile. For the negatively biased case, the opposite is true. Figure \ref{fig:phi-ve} shows that experimental $V_E$ flow is negative for $R < 1.2$ m. In this case, the bottom is the downstream side and the bottom density profile is higher than the top density profile. However, for the values of vertical \ExB\ flow given in Fig.~\ref{fig:phi-ve}b and also in Ref.~\citenum{gentle2014turbulence}, this toy model predicts a downstream density that is 2--3 times that of the upstream density, which is not observed. This model neglects effects such as drag on ion flow due to neutral interactions, finite ion temperature and variability of electron temperature. Thus, we do not anticipate strong quantitative agreement with this model. 

\begin{figure}
    \centering
    \includegraphics[width=.5\textwidth]{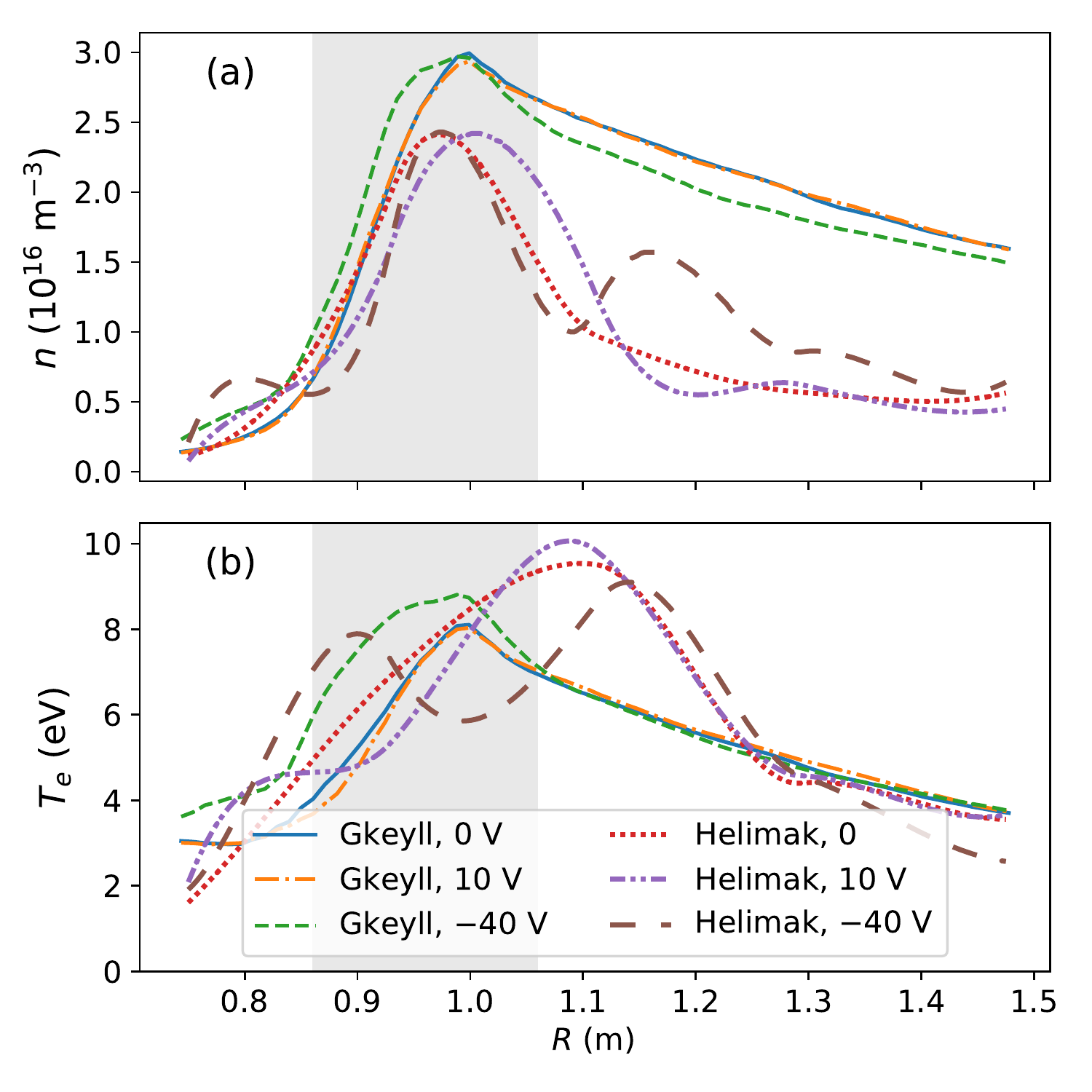}
    \caption{Equilibrium profile comparison of grounded and limiter-biased scenarios. Plot (a) contains electron density profiles for simulation and experiment and (b) contains electron temperature profiles. Experimental profiles were measured at the bottom of the device. All profiles were calculated by averaging in time, and also in $y$ for simulation data.}
    \label{fig:eqprof-bias}
\end{figure}

Fig.~\ref{fig:eqprof-bias}a compares electron density equilibrium profiles from simulation and experiment for the grounded and biased scenarios. Profiles were calculated at the $\zmin$ end of the domain and compared with corresponding experimental profiles measured at the bottom of the Helimak. Since the vertical \ExB\ flow is not included in the simulation, equilibrium density profiles are virtually identical at the top and bottom of the $z$-domain. This may also explain why there is little difference in simulation density profiles among the three cases. By comparison, experimental profiles exhibit differences in both magnitudes and gradients.

Fig.~\ref{fig:eqprof-bias}b contains a similar comparison of equilibrium electron temperature profiles. The simulation produced a slight increase in the temperature profile within the biased region and just to the left of it. This is consistent with the fact that a more negative bias voltage produces a larger potential well that reflects more high-energy electrons than in the grounded or positively biased case. The profiles from the latter two are very similar. Likewise, the experimental temperature profile from the negatively biased case is significantly different than the other two experimental profiles. Differences between experiment and simulation temperature profiles is likely due to the fact that most of the power is absorbed at the upper hybrid resonance, which depends on plasma density. Therefore, changes in the density profile lead to changes in power absorption and, in turn, equilibrium temperature profiles. This density dependence is not currently contained within the \gkyl\ source model.

 \begin{figure*}
    \centering
    \includegraphics[width=.9\textwidth]{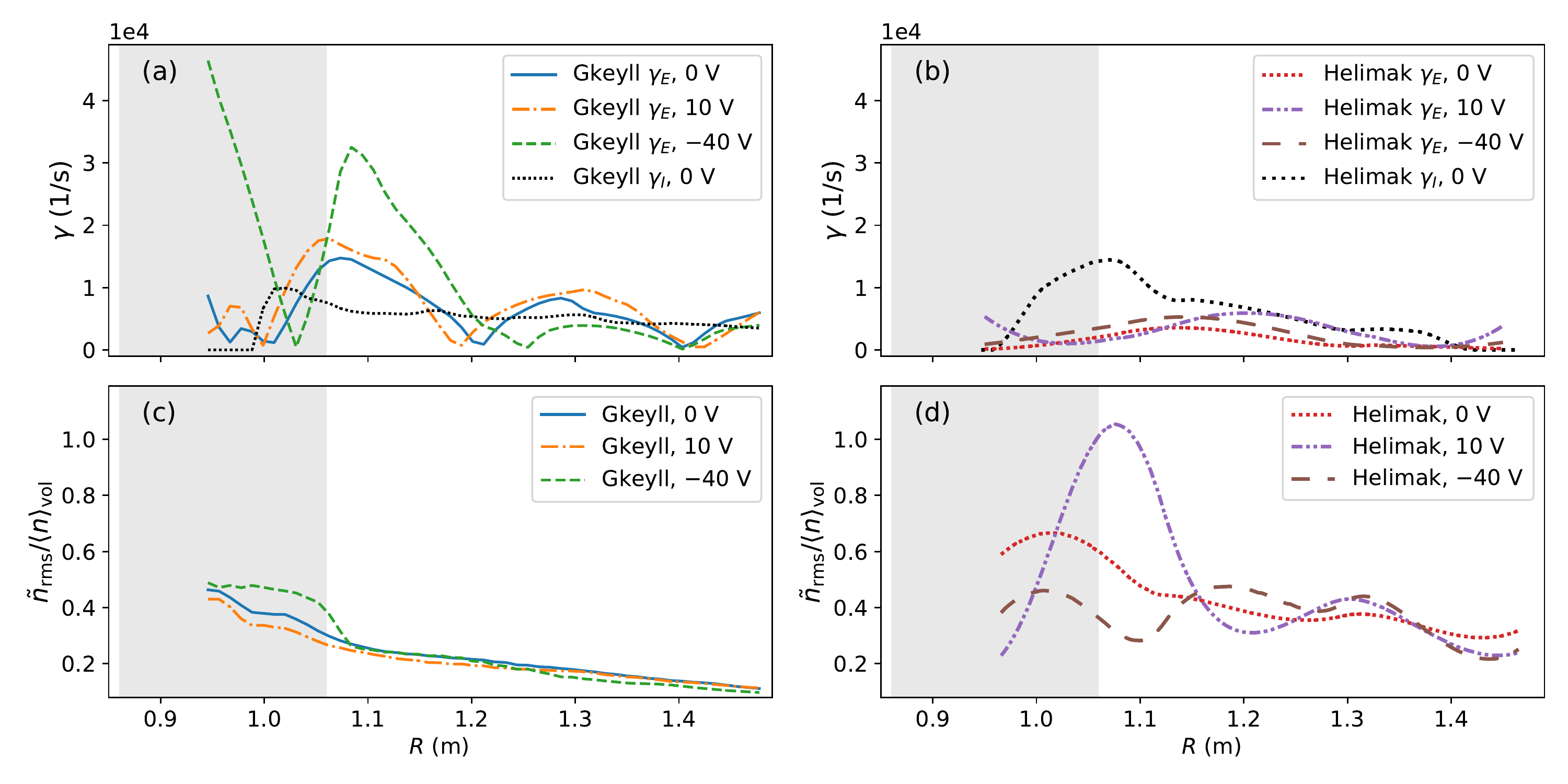}
    \caption{Plots (a) and (b) compare the magnitude of the \ExB\ shear, $\gamma_E = |dV_E/dx|$, with the grounded local interchange growth rate $\gamma_I = c_s/\sqrt{R L_n}$ from \gkyl\ simulations and the Helimak experiment, respectively. Plots (c) and (d) compare the root mean square density fluctuation levels for the grounded and limiter-biased scenarios of the simulation and experiment, respectively. Density fluctuation levels are normalized to the volume-averaged density $\langle n \rangle_{\rm vol}$.}
    \label{fig:shear-dn}
\end{figure*}

\subsection{Comparison of turbulence levels}
According to shear-suppression theory, the shear rate should be larger than the decorrelation rate or instability growth rate in order for shear to break up turbulent eddies.\cite{terry2000suppression} Therefore, we compare \ExB\ shear, $\gamma_E = |dV_E/dx|$, of the three scenarios to the linear interchange growth rate from the grounded scenario in Fig.~\ref{fig:shear-dn}a and Fig.~\ref{fig:shear-dn}b. The local, linear interchange growth rate for the grounded scenario is calculated as $\gamma_I = c_s /\sqrt{R L_n}$ for $L_n > 0$ and $\gamma_I = 0$ for $L_n \leq 0$. The density gradient scale length is $L_n = -n/(dn/dx)$, using the grounded density equilibrium profiles from Fig.~\ref{fig:eqprof-bias}a. We focus on the bad-curvature region ($R > 1.0$), where $\gamma_I > 0$. In Fig.~\ref{fig:shear-dn}a, simulation shear rates are only marginally larger than the simulation interchange growth rate, except for the negatively biased case, in which the shearing rate is approximately twice as large near $R=1.1$ m. In Fig.~\ref{fig:shear-dn}c, the root mean square (RMS) of density fluctuations from the simulation have been normalized to a volume-averaged density, and these plots show no significant changes in turbulence levels between the biased and unbiased cases in the bad-curvature region. Given the large peaks in the shear rate of the negatively biased scenario, one might expect a corresponding reduction in turbulence levels. However, there is actually a slight increase in this case. These differences are marginal, and it is not possible to draw further conclusions on the effect of shear-suppression of turbulence in the simulations. In Fig.~\ref{fig:shear-dn}b, experimental shear levels are less than the local interchange growth rate obtained from the grounded experimental equilibrium density profile. From this comparison, one might expect little changes in the density turbulence fluctuation levels. However,  Fig.~\ref{fig:shear-dn}d, shows an increase in turbulence fluctuations for the positively biased case and a small reduction in the negatively biased case. 

\begin{figure}
    \centering
    \includegraphics[width=.5\textwidth]{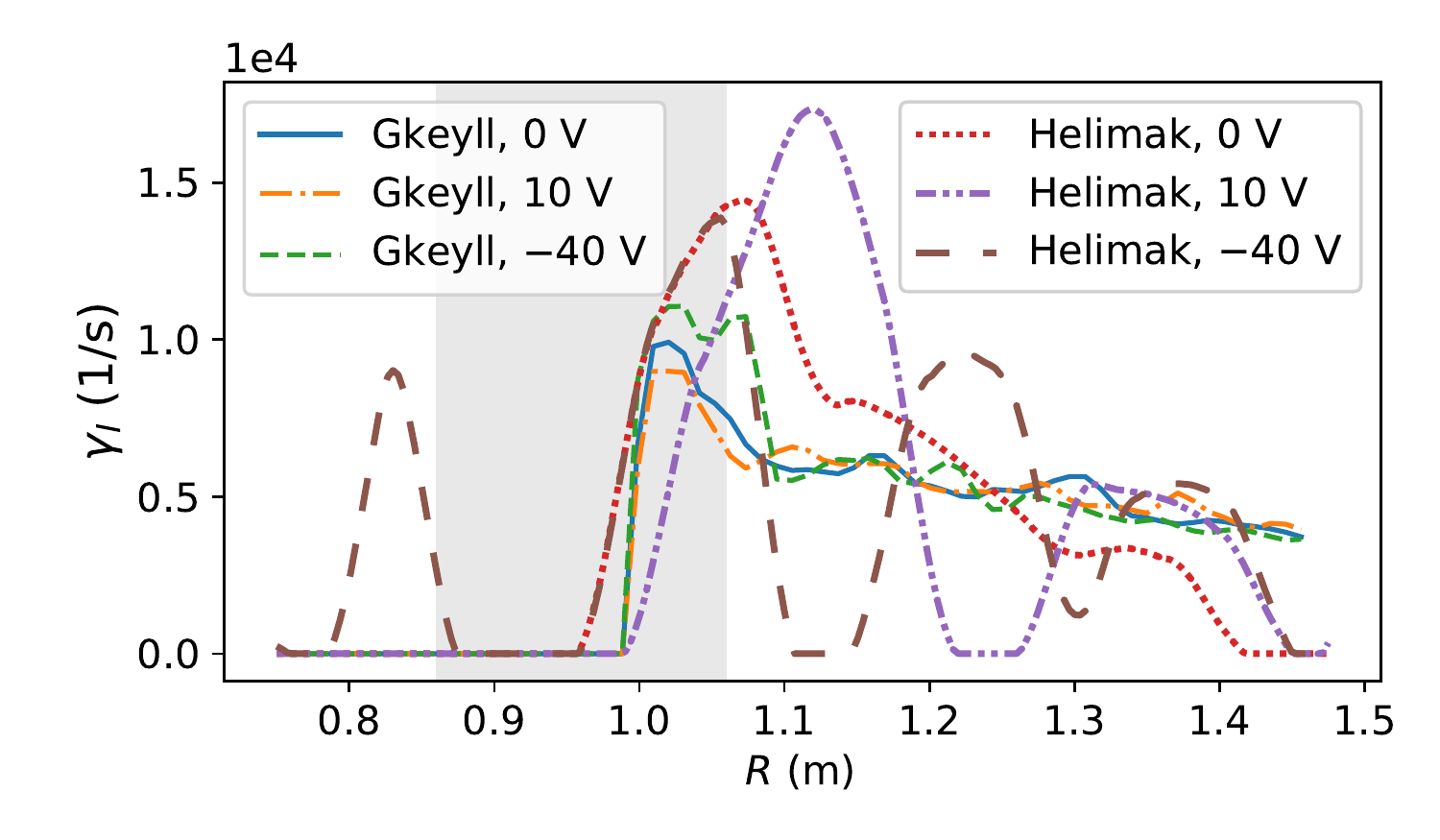}
    \caption{Local interchange growth rates for the grounded and limiter-biased cases of the Gkeyll simulation and Helimak experiment are compared.}
    \label{fig:gammaI}
\end{figure}

A comparison of the local, linear interchange growth rates for grounded and biased scenarios may explain the differences and similarities of turbulence profiles. These are shown in Fig.~\ref{fig:gammaI}. The interchange growth rates from simulation are very similar except for an increase in $\gamma_I$ at the right edge of the biased region for the negatively biased scenario between 1.0 and 1.1 m. This corresponds to a slight increase in turbulence levels seen in Fig.~\ref{fig:shear-dn}c. Experimental growth rate values from the grounded and biased scenarios differ more significantly from one another, and relative increases or decreases in these values correspond to relative increases or decreases in the turbulence levels of Fig.~\ref{fig:shear-dn}d. Thus, it appears that bulk transport due flow shear may change equilibrium profile gradients, which in turn may affect turbulence levels.

\begin{figure}
    \centering
    \includegraphics[width=0.5\textwidth]{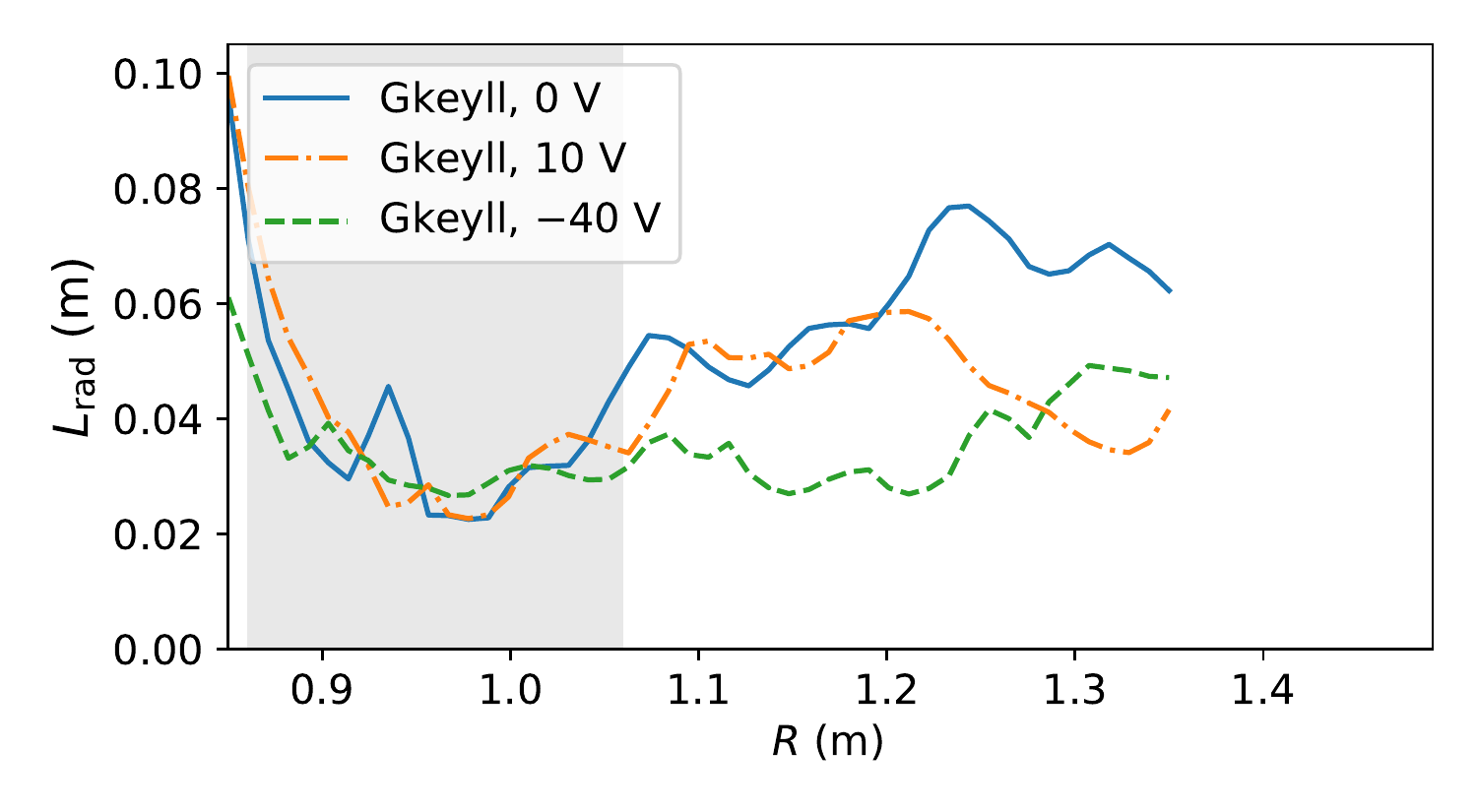}
    \caption{Radial correlation lengths, $L_{rad}$, for the grounded and limiter-biased cases of the simulations are compared.}
    \label{fig:lrad-bias}
\end{figure}
Radial correlation lengths can be an indication of the size of turbulent structures in a plasma, and these are plotted in Fig.~\ref{fig:lrad-bias} for the simulation scenarios. There is a small reduction in radial correlation lengths for the negatively biased case at radii greater than 1.1 m. A peak in the shear rate occurs at 1.1 m in Fig.~\ref{fig:shear-dn}a, suggesting that the stronger shear in this region could be breaking up turbulent structures and reducing overall correlation lengths. On the other hand, shearing rates for the positively biased case are not significantly higher than the grounded scenario, and we are not able to draw conclusions about the decrease in radial correlation lengths observed around 1.2 m. Experimental radial correlation length data was not available for this comparison, though previous measurements give radial correlation lengths that are about 0.04 m or less.\cite{gentle2014turbulence} Thus, the simulation over-predicts these values.

\begin{figure}
    \centering
    \includegraphics[width=.5\textwidth]{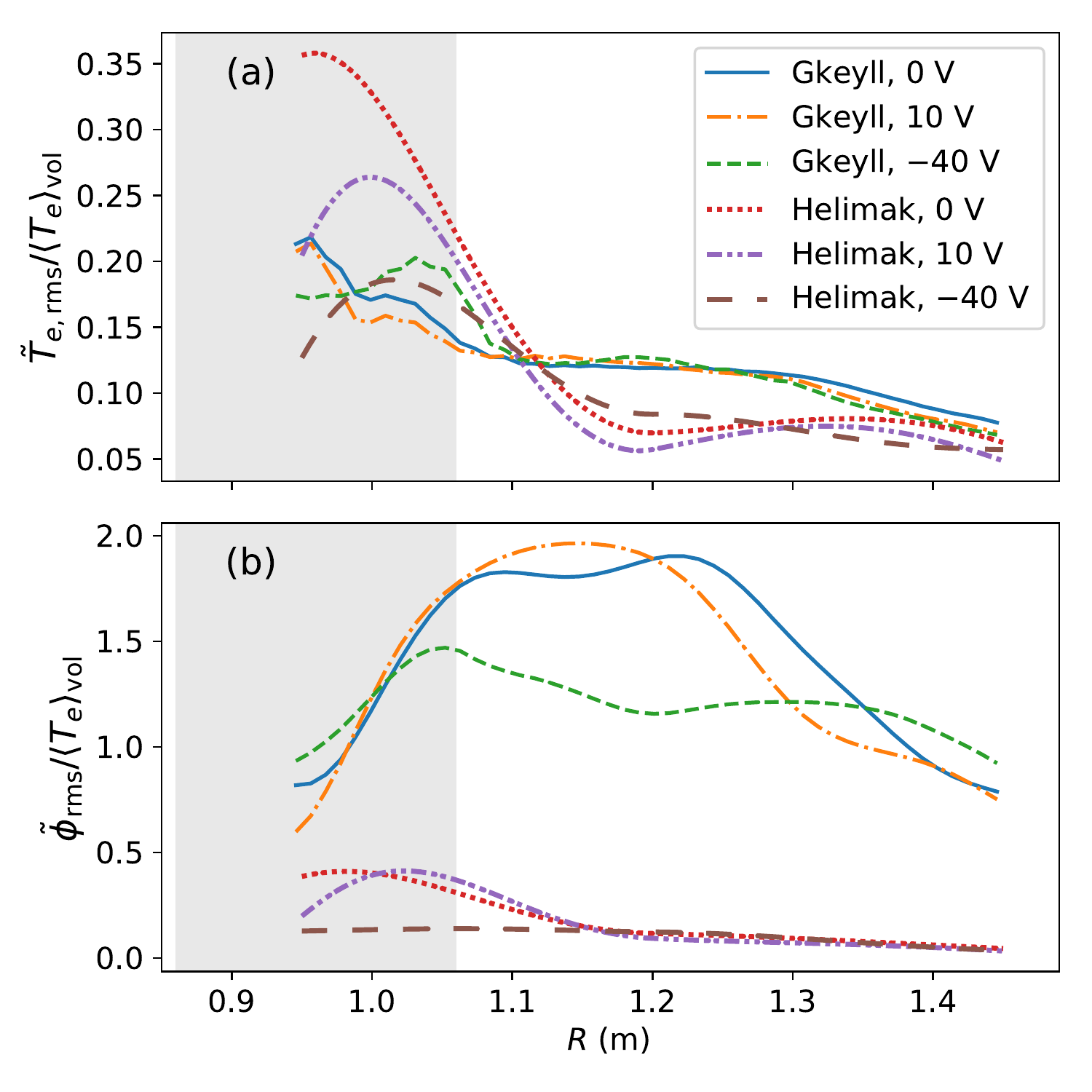}
    \caption[Comparisons of electron temperature fluctuations and potential fluctuations for the grounded and limiter-biased scenarios of simulation and experiment.]{In (a) comparisons of electron temperature fluctuations and in (b) potential fluctuations for the grounded and limiter-biased scenarios. Both are normalized to the volume-averaged electron temperature $\langle T_e \rangle_{\rm vol}$.}
    \label{fig:dTdPhi}
\end{figure}

\subsection{Other measurements}
Fluctuations in electron temperature and plasma potential were also measured. Figure \ref{fig:dTdPhi}a contains radial profiles of electron temperature fluctuations for simulation and experiment. This is calculated as the RMS of temperature fluctuations normalized to the volume-averaged electron temperature. The simulation data shows little change between the grounded and positively biased cases, and a more noticeable change in the profile of the negatively biased case. For the latter, the fluctuations decrease at the center of the biased region and increase near the edges. The experimental data displays a decrease in fluctuations for both biased cases, as compared with the grounded scenario. Figure \ref{fig:dTdPhi}b contains radial profiles of plasma potential fluctuations for simulation and experiment. Again the RMS values of potential fluctuations are normalized to the volume-averaged electron temperature. The simulation fluctuation values are up to five times larger than experimental values. Both simulation and experimental data have grounded and positively biased cases that are very similar to one another, and the corresponding negatively biased case is reduced to the right of the biased region. This relative decrease is more significant in the experimental data. 

\begin{figure}
    \centering
    \includegraphics[width=.5\textwidth]{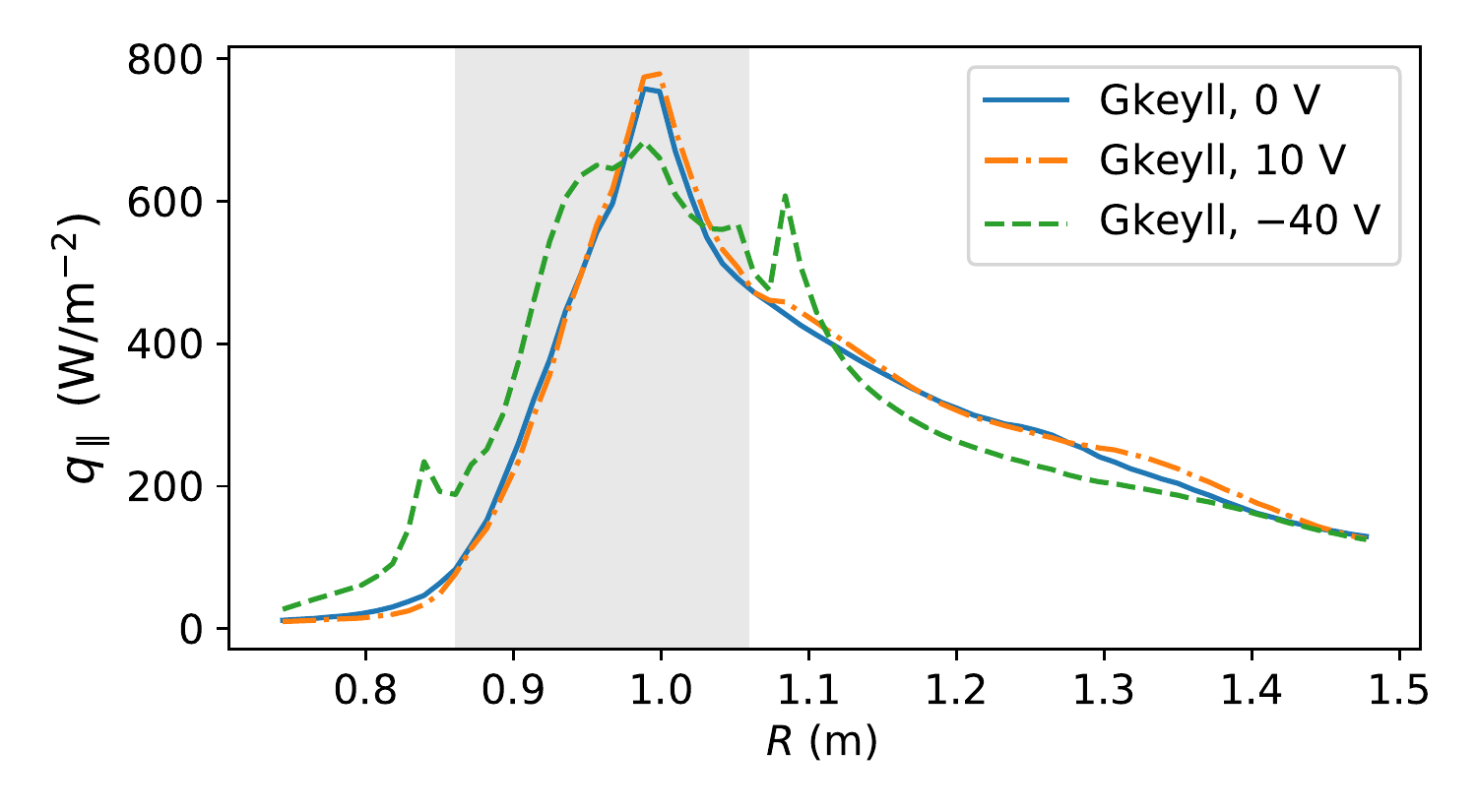}
    \caption{Comparison of parallel heat flux to top plates from grounded and limiter-biased simulation scenarios.}
    \label{fig:q-bias}
\end{figure}
Figure \ref{fig:q-bias} compares the heat flux to the conducting plates from grounded and limiter-biased simulations. This was calculated as $q_\parallel = \langle q_{\parallel,e} \rangle + \langle q_{\parallel,i} \rangle$, where $\langle ... \rangle$ denotes an average in time and in $y$. The electron and ion heat fluxes are given by
\begin{eqnarray}
    q_{\parallel,e} = \int_{v_c}^{\vparmax} d^3 v \; f_e H_e \vpar \\
    q_{\parallel,i} = \int_{0}^{\vparmax} d^3 v \; f_i H_i \vpar ,
\end{eqnarray}
where $v_c$ is the electron parallel cutoff velocity and $f_s$ and $H_s$ are evaluated at $z=\zmax$. The grounded and positively biased cases are very similar. The peak of the negatively biased scenario is slightly reduced, and the $x$-integrated value of the heat flux is less than the other cases. This is consistent with the higher simulated electron temperature profile from visible in Fig.~\ref{fig:eqprof-bias}b. Heat flux data from the Helimak experiment was not available.

\section{Conclusions} \label{sec:conc}
In this paper, we presented simulations of limiter biasing in the Texas Helimak and compared with grounded simulations and with experimental data. Limiter biasing is applied in the Helimak to better understand the role that \ExB\ flow shear can play on turbulence suppression. The limiter biasing was simulated self-consistently through the conducting sheath boundary conditions outlined in Sec.~\ref{sec:modeleqns}. As in Ref.~\citenum{bernard2019}, we observe a top--bottom asymmetry in experimental density profiles that is not present in our simulation. We have demonstrated that this is likely due to the vertical (or analogously to tokamak geometry, poloidal) component of the \ExB\ flow that is not currently included in our model. This vertical (or poloidal) \ExB\ flow is commonly neglected in flux-tube codes, but our results demonstrate the shortcomings of this assumption. By a simplified 1D transport model (Appendix \ref{app}), we demonstrate that a moderate vertical \ExB\ flow, which is of the same order of magnitude as the vertical component of the parallel flow, may cause an increase in density levels at the downstream end of the device, towards which the \ExB\ flow is directed. A downstream increase is observed in experimental density profiles.

Our analysis shows that local linear interchange growth rates are generally larger or only marginally less than flow shear rates, particularly in the experimental data, which suggests that shear would not have a significant effect on turbulence levels. By comparing the local linear interchange growth rates for the biased scenarios, we observed that an increase or decrease relative to the grounded scenario corresponded to an increase or decrease, respectively, in turbulence levels. We thus conclude that differences in experimental turbulent fluctuation levels observed for different bias voltages in the Helimak likely resulted from changes in equilibrium profile gradients caused by changes in bulk flow. In agreement with Ref.~\citenum{gentle2014turbulence}, we found no strong evidence for shear-suppression of turbulence.

To more accurately capture the top--bottom asymmetry and changes to turbulence levels observed experimentally it is necessary to implement realistic geometry in our model that includes the vertical \ExB\ flow and magnetic shear. These developments are currently underway and validation of these features will be forthcoming. A more realistic model would also include interaction with neutrals that can produce a drag force on ions and affect flow patterns. This may account for differences in the vertical flow calculated from spectroscopic measurements presented in Fig.~14 of Ref.~\citenum{gentle2014turbulence} as compared with the \ExB\ flow calculated from the baffled probe measurements of the plasma potential shown in Fig.~\ref{fig:phi-ve}. Thus coupling a neutral transport model to the \gkyl\ gyrokinetic solver is important future work.

\begin{acknowledgments}
We thank J.~Guterl and F.~D.~Halpern for helpful feedback and enlightening discussions. This material is based upon work supported by the U.S. Department of Energy, Office of Science, Office of Fusion Energy Sciences, Theory Program, under Award No.~DE-FG02-95ER54309; by DOE contract DE-FG02-04ER-54742, through the Institute of Fusion Studies at the University of Texas at Austin; and by DOE contract DE-AC02-09CH11466, through the Princeton Plasma Physics Laboratory. This work used the Extreme Science and Engineering Discovery Environment (XSEDE), which is supported by National Science Foundation grant number ACI-1548562.

-- Disclaimer -- This report was prepared as an account of work sponsored by an agency of the United States Government. Neither the United States Government nor any agency thereof, nor any of their employees, makes any warranty, express or implied, or assumes any legal liability or responsibility for the accuracy, completeness, or usefulness of any information, apparatus, product, or process disclosed, or represents that its use would not infringe privately owned rights. Reference herein to any specific commercial product, process, or service by trade name, trademark, manufacturer, or otherwise, does not necessarily constitute or imply its endorsement, recommendation, or favoring by the United States Government or any agency thereof. The views and opinions of authors expressed herein do not necessarily state or reflect those of the United States Government or any agency thereof.
\end{acknowledgments}

\appendix

\section{1D transport model for vertical $E \times B$ flow} \label{app}
We now set forth a simplified toy model to help explain the top--bottom asymmetry in density profiles observed in the Helimak. We begin with the zeroth and first moments of the 6D kinetic Boltzmanm equation for a single plasma species:
\begin{eqnarray}
    \pd{n_s}{t} + \nabla \cdot (\v{u}_s n_s) &=& S_s \\
    \pd{m_s n_s \v{u}_s}{t} + \nabla \cdot (m_s n_s \v{u}_s \v{u}_s) &=& n_s Z_s e (\v{E} + \v{u}_s \times \v{B}) \nonumber \\ &-& \nabla p_s + \v{F}_s, \label{eq:mom1}
\end{eqnarray}
where we have assumed an isotropic pressure $p_s$. We also assume the external force $\v{F}_s$ is zero. Using a simple cylindrical geometry $(r,\varphi,z)$, the direction of the helical magnetic field becomes $\bhat \doteq \v{B}/B = b_\varphi\uv{\varphi} + b_z\uv{z}$. $B$, $b_\varphi$, and $b_z$ are assumed to be only functions of $r$. We assume that any ionized neutrals are cold, i.e. they are born with zero velocity. Hence, ionization is a source of density but not momentum. 

We construct a 1D transport equation by only considering flows within a constant $r$ surface. Thus the radial component of the fluid flow is zero ($u_{s,r} =0$). We assume that all plasma quantities and the plasma potential are time-independent and axisymmetric. First, consider the conservation of parallel momentum by dotting Eq.~\ref{eq:mom1} with $\bhat$, and consider just the second term. Using the definition for the divergence of a tensor in cylindrical coordinates and assuming axisymmetry gives
\begin{eqnarray}
    \bhat \cdot \nabla \cdot (m_s n_s \v{u}_s \v{u}_s) &=& b_\varphi \partial_z(m_s n_s u_{\varphi s} u_{zs}) 
    + b_z \partial_z (m_s n_s u_{zs} u_{zs}) \nonumber \\
    &=& \partial_z (m_s n_s \upars u_{zs}),
\end{eqnarray}
where $\upars \doteq \bhat \cdot \v{u}_s$. With the definitions $\nabla_\parallel \doteq \bhat \cdot \nabla$, $E_\parallel \doteq \bhat \cdot \v{E}$, we can construct our system of equations for the 1D model:
\begin{eqnarray}
    \partial_z (n_s u_{sz}) = S_s \\
    \partial_z (m_s n_s \upars u_{zs}) = n_s Z_s e E_\parallel - \nabla_\parallel p_s. \label{eq:mom1par}
\end{eqnarray}
For the purposes of this model, let us assume that the vertical flow is given by $u_{sz} = b_z\upars + u_{Ez}$, where $u_{Ez}$ is the vertical component of the \ExB\ flow and that temperature is constant. The momentum equation becomes
\begin{equation}
    \partial_z(m_s n_s \upars u_{Ez} + m_s b_z n_s \upars^2) + n_s Z e \nabla_\parallel \phi = -T_s \nabla_\parallel n_s.
\end{equation}
We will consider a case with $u_{Ez} \sim b_z c_s$. To further reduce the model, assume ions are cold and that electron mass is negligible. The electron equation can be rearranged to give
\begin{equation}
    \nabla_\parallel \phi = \frac{T_e}{e}\nabla_\parallel \ln n_e,
\end{equation}
while the ion equations are 
\begin{eqnarray}
    \partial_z (n_i u_{zi}) = S_i \\
    \partial_z (m_i n_i \upari u_{zi}) = - n_i Z_i e \nabla_\parallel \phi
\end{eqnarray}
Assuming quasineutrality gives $n_e = Z_i n_i$ and thus $\nabla_\parallel \ln n_e = \nabla_\parallel \ln Z_i n_i = \nabla_\parallel \ln n_i$. Assuming axisymmetry gives $\nabla_\parallel = b_z \partial_z$, and the ion momentum equation becomes
\begin{equation}
    \partial_z (m_i n_i \upari u_{zi}) = - b_z \partial_z (Z_i T_e n_i).
\end{equation}
We can also assume $b_z$ is independent of $z$ as in the Helimak to rewrite the momentum equation again
\begin{equation}
    \partial_z [n_i (m_i \upari u_{iz} + b_z Z_i T_e)] = 0,
\end{equation}
which indicates that the quantity in brackets is independent of $z$. This is a generalization of the kinetic pressure balance. 

Let us assume that the source $S_i$ is known so that we can define the vertical particle flux up to some constant:
\begin{equation}
    n_i u_{iz} = \Gamma(z) \doteq \Gamma_0 + \zintm dz' \, S_i.
\end{equation}
For a positive source $S_i$, the flux $\Gamma$ should be a monotonically increasing function of $z$. Integrating the momentum equation and making appropriate substitutions using $\upari = (u_{iz} - u_{Ez})/b_z$ and $u_{iz} = \Gamma/n_i$ gives
\begin{eqnarray}
    \Pi_0 &=& n_i (m_i \upari u_{iz} + b_z Z_i T_e) \nonumber \\
    &=& \frac{\Gamma m_i}{b_z} \left( \frac{\Gamma}{n_i} - u_{Ez} \right) + n_i b_z Z_i T_e \\
    0 &=& b_z^2 Z_i T_e n_i^2 - \left(\Gamma m_i u_{Ez} + \Pi_0 b_z \right) n_i + \Gamma^2 m_i. \label{eq:M1int}
\end{eqnarray}
Note that this equation can also be written in terms of $u_{iz}$, though we will continue our analysis in terms of $n_i$. Note that there is no explicit $z$-dependence in these equations and $\Gamma$ is taken to be the independent variable. The final solution can thus be mapped back to $z$, since we have assumed $\Gamma$ is known in terms of $z$.

Equation (\ref{eq:M1int}) can be solved with the quadratic formula:
\begin{equation}
    n_i = \frac{\Pi_0 b_z + \Gamma u_{Ez} m_i \pm 
    \sqrt{(\Pi_0 b_z + \Gamma u_{Ez} m_i)^2 - 4b_z^2Z_i T_e \Gamma^2 m_i}}{2 b_z^2 Z_i T_e}. \label{eq:M1dens}
\end{equation}
We choose the sign on the radical so that the density remains nonzero when $\Gamma = 0$. We define $n_{i0}$ to be the density at this stagnation point:
\begin{equation}
    n_{i0} \doteq \frac{\Pi_0}{b_z Z_i T_e} > 0.
\end{equation}
We now perform the following normalizations: $\nbar \doteq n_i/n_{i0}$, ${\ubar \doteq u_{Ez}/|b_z|c_s}$, and $\Gbar \doteq \Gamma/n_{i0}|b_z|c_s$, where $c_s^2 = Z_i T_e/m_i$. Then Eq.~(\ref{eq:M1dens}) becomes simply
\begin{equation}
    \nbar^2 - (1 + \Gbar\ubar)\nbar + \Gbar^2 = 0.
\end{equation}
The density can now be solved in terms of the normalized flux, which is the new independent variable, giving
\begin{equation}
    \nbar = \frac{1}{2} \left[ 1 + \Gbar\ubar + \sqrt{(1 + \Gbar\ubar)^2 - 4 \Gbar^2}\right]. \label{eq:nbar}
\end{equation}
First, consider the case with negligible vertical \ExB\ flow. The normalized density equation reduces to ${\nbar = \frac{1}{2}\left[ 1 + \sqrt{1 - 4\Gbar^2}\right]}$. The normalized density reaches its maximal value of 1 at $\Gbar = 0$ and reaches a minimal value of $1/2$ when $|\Gbar|= 1/2$. This means the maximum dimensional flow is 
\begin{equation}
    |u_{iz}| = \frac{|\Gamma|}{n_i} = |b_z|c_s\frac{|\Gbar|}{\nbar} \leq |b_z|c_s,
\end{equation}
and $|\upari| \leq c_s$. Combined with the Bohm sheath requirement of $|\upari| \geq c_s$, this implies that 
\begin{equation}
n_{i0} |b_z|c_s = \zintmm dz' S_i.
\end{equation}
(Note that this is the same result given by the 1D transport model in the Appendix of Ref.~\citenum{shi2019full}.)

\begin{figure}
     \centering
     \includegraphics[width=.5\textwidth]{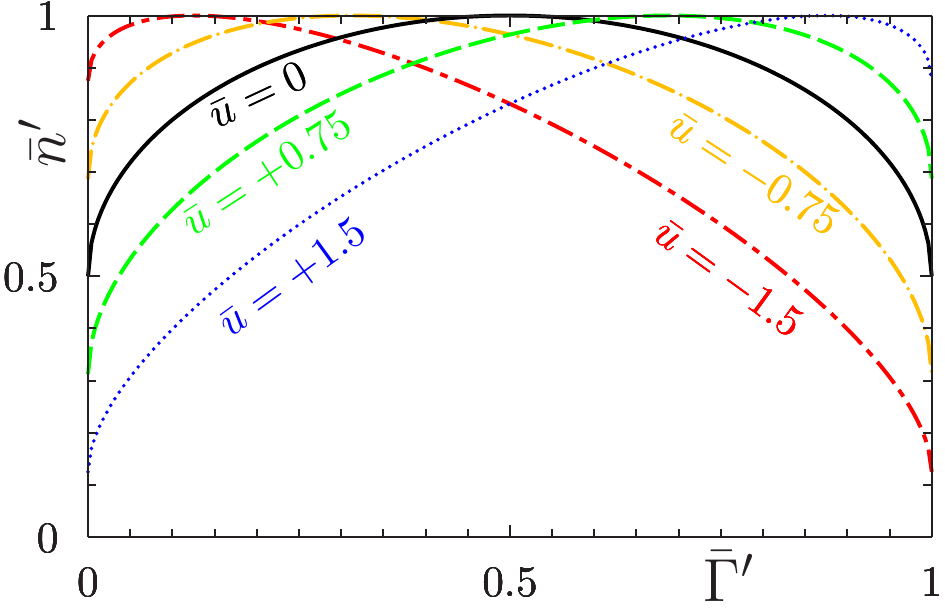}
     \caption{Normalized density $\nbar'$ as a function of normalized $z$ position $\Gbar'$ for normalized, vertical \ExB\ flow $\ubar = -1.5$, $-0.75$, $0$, $0.75$, and $1.5$ demonstrates asymmetry in the density profile for nonzero $\ubar$.}
     \label{fig:n_of_Gamma}
 \end{figure}

Now consider a nonzero ``moderate'' \ExB\ flow, $|\ubar| \leq 2$. Equation (\ref{eq:nbar}) represents a real solution when $(1 + \Gbar\ubar)^2 \geq 4\Gbar^2$, thus for
\begin{equation}
    -\frac{1}{2 + \ubar} \leq \Gbar \leq \frac{1}{2 - \ubar}.
\end{equation}
Evaluating Eq.~(\ref{eq:nbar}) at the lower and upper bounds yields
\begin{eqnarray}
    \nbar_- &\doteq& \nbar\left(\Gbar = -1/(2 + \ubar)\right) = 1/(2 + \ubar) \\
    \nbar_+ &\doteq& \nbar\left(\Gbar =  1/(2 - \ubar)\right) = 1/(2 - \ubar),
\end{eqnarray}
so that these boundaries again correspond to poloidal-sonic outflow, $|u_{iz}|
 = |b_z|c_s$. If we assume that the sheath requires outflow at least this strong,\cite{Rozhansky1994} then the domain in $\Gbar$ is $[-(2 + \ubar)^{-1}, (2 - \ubar)^{-1}]$, which lets us relate
 \begin{eqnarray}
     n_{i0}|b_z|c_s/(1-\ubar^2/4) &=& \Gamma(\zmax) - \Gamma(\zmin) \nonumber \\
     &=& \zintmm dz' S_i \doteq S_{i, \rm tot}.
 \end{eqnarray}
 Using this, we can define new dimensionless variables with $\ubar$-independent normalizations:
 \begin{eqnarray}
     \Gbar' \doteq S^{-1}_{i, \rm tot} \zintm dz' \, S_i = \frac{\Gbar(z) - \Gbar(\zmin)}{\Gbar(\zmax) - \Gbar(\zmin)} \nonumber \\
     = \frac{1}{4} [(4 - \ubar^2)\Gbar(z) + 2 - \ubar], \\
     \nbar' \doteq \frac{n_i}{S_{i, \rm tot}/|b_z|c_s} = (1 - \ubar^2/4)\nbar,
 \end{eqnarray}
 in terms of which the density profiles can be written
 \begin{equation}
     \nbar' = \frac{1}{4} \left[ 2 \Gbar' \ubar + 2 - \ubar + 2 \sqrt{\Gbar'(1 - \Gbar')(4 - \ubar^2)}\right],
 \end{equation}
 as plotted in Fig.~\ref{fig:n_of_Gamma}. The new domain is $0 \leq \Gbar' \leq 1$, so we can evaluate the density asymmetry for a given $\ubar$ as 
 \begin{equation}
     \nbar'(\Gbar' = \frac{1}{2} + s) - \nbar'(\Gbar' = \frac{1}{2} - s) = s\ubar,
 \end{equation}
 where $0 \leq s \leq 1/2$, and $\nbar'$ is larger on the downstream side as anticipated.
 
 This simple toy model provides a potential explanation for the top--bottom asymmetry observed in the Helimak density profiles, in qualitative agreement with experimental data. However, the toy model is extremely simple, neglecting not only ion temperature and the effects of neutral interactions, but also constraints on the $z$-integrated parallel potential variation and self-consistency between $\nabla_\parallel \phi$ and the \ExB\ drift, which would require a 2D treatment. We, therefore, do not expect strong quantitative agreement with experimental observations.

\bibliography{references}

\end{document}